\documentclass[superscriptaddress, reprint]{revtex4-2}
\usepackage{multirow}
\usepackage{hyperref}
\usepackage[version=4]{mhchem} 
\usepackage{graphicx}
\usepackage{chemmacros}
\usepackage{gensymb}
\usepackage{dcolumn}
\usepackage{amssymb}

\newcommand{\Ad}{A }
\newcommand{\Add}{A{\textquotesingle}}
\newcommand{\tac}{A$_{2}$A{\textquotesingle}A{\textquotesingle}{\textquotesingle}B$_4$O$_{12}$ }
\newcommand{\tacm}{A$_{2}$A{\textquotesingle}A{\textquotesingle}{\textquotesingle}Mn$_4$O$_{12}$ }

\newcommand{\Ypn}{$x=0.23$ }
\newcommand{\Ypnf}{$x=0.16$ }
\begin{document}    

\title{Magnetic inhomogeneities in the quadruple perovskite manganite \ce{[Y_{2-x}Mn_x]MnMnMn4O12}}

\author{A. M. Vibhakar}
\affiliation{Diamond Light Source Ltd,  Harwell Science and Innovation Campus, Didcot, Oxfordshire, OX11 0DE, United Kingdom}
\author{D. D. Khalyavin}
\affiliation{ISIS facility, Rutherford Appleton Laboratory-STFC, Chilton, Didcot, OX11 0QX, United Kingdom}
\author{P. Manuel}
\affiliation{ISIS facility, Rutherford Appleton Laboratory-STFC, Chilton, Didcot, OX11 0QX, United Kingdom}
\author{N. J. Steinke}
\affiliation{Institut Laue-Langevin, 71 avenue des Martyrs, F-38042 Grenoble, France}
\author{L. Zhang}
\affiliation{International Center for Materials Nanoarchitectonics (WPI-MANA), National Institute for Materials Science (NIMS), Namiki 1-1, Tsukuba, Ibaraki 305-0044, Japan}
\affiliation{Graduate School of Chemical Sciences and Engineering, Hokkaido University, North 10 West 8, Kita-ku, Sapporo, Hokkaido 060-0810, Japan}
\author{K. Yamaura}
\affiliation{International Center for Materials Nanoarchitectonics (WPI-MANA), National Institute for Materials Science (NIMS), Namiki 1-1, Tsukuba, Ibaraki 305-0044, Japan}
\affiliation{Graduate School of Chemical Sciences and Engineering, Hokkaido University, North 10 West 8, Kita-ku, Sapporo, Hokkaido 060-0810, Japan}
\author{A. A. Belik}
\affiliation{International Center for Materials Nanoarchitectonics (WPI-MANA), National Institute for Materials Science (NIMS), Namiki 1-1, Tsukuba, Ibaraki 305-0044, Japan}
\author{R. D. Johnson}
\affiliation{Department of Physics and Astronomy, University College London, Gower Street, London, WC1E 6BT, United Kingdom}
\date{\today}

\begin{abstract}
A combination of competing exchange interactions and substitutional disorder gives rise to magnetic inhomogeneities in the \ce{[Y_{2-x}Mn_x]MnMnMn4O12} $x = 0.23$ and $x = 0.16$ quadruple perovskite manganites. Our neutron powder scattering measurements show that both the $x = 0.23$ and $x = 0.16$ samples separate into two distinct magnetic phases; below T$_{1}$ = 120 $\pm$ 10 K the system undergoes a transition from a paramagnetic phase to a phase characterised by short range antiferromagnetic clusters contained in a paramagnetic matrix, and below T$_{2}$ $\sim$ 65 K, the system is composed of well correlated long range collinear ferrimagnetic order, punctuated by short range antiferromagnetic clusters. A sharp increase in the antiferromagnetic phase fraction is observed below $\sim$ 33 K, concomitant with a decrease in the ferrimagnetic phase fraction. Our results demonstrate that the theoretically proposed AFM phase is stabilised in the \ce{[Y_{2-x}Mn_x]MnMnMn4O12} manganites in the presence of dominant B-B exchange interactions, as predicted. 
\end{abstract}

\maketitle
Competing exchange interactions and disorder are two parameters that can be tuned to access a wide range of macroscopically distinct magnetic phases. For instance, disordered magnetic phases such as spin glasses arise from the presence of strong quenched disorder and competing exchange interactions. While unordered magnetic systems, such as spin liquids, typically arise from strong competing exchange and occur in the absence of quenched disorder. Competing exchange and disorder also form essential ingredients for realising Griffiths phases \cite{2006Magen}, which occur in magnetic systems when clusters of magnetic order form that are large enough to produce a singularity in the free energy \cite{1969Griffiths, 1982Bray, 1987Bray}. Imaging and understanding the mathematical properties of such magnetic phases has shaped our understanding of neural networks \cite{1990Anderson} and complex matter \cite{2005Dagotto}, and continues to play an important role in demonstrating novel forms of complex behaviour \cite{2005Dagotto}. 

An exemplary family of materials that demonstrate the interplay between disorder and competing exchange are the simple perovskite manganites. In the absence of any substitutional disorder, i.e for A$^{3+}$MnO$_3$, well ordered antiferromagnetic (AFM) ground state structures are stabilised. The introduction of hole doping, through the substitution of a divalent cation on the A sites, gives rise to an electronic texture that is inhomogeneous on the nanoscale. In general for \ce{A(III)_{1-x}A{\textquotesingle}(II)_{x}MnO3} x $<$ 0.5, an insulator to metal phase transition is observed below T$_\mathrm{C}$, and the system is observed to order ferromagnetically in the metallic phase owing to the double exchange mechanism \cite{1951Zener}. Numerous studies have found that there is a finite temperature interval that precedes the phase transition to ferromagnetic (FM) order, where the system is characterised by coexistence of paramagnetic insulating and FM metallic regions, a Griffiths phase \cite{1996Lynn, 1997Teresa}. Here, charge doping causes the separation of these systems into the two distinct electronic phases, giving rise to the two magnetic phases and the establishment of a Griffiths phase. Furthermore for certain compositions of the mixed valence manganites \cite{1999Uehara}, such as in \ce{Pr_{0.7}Ca_{0.3}MnO3} \cite{1998Cox}, the randomly dispersed Mn$^{4+}$ on the B-site sublattice give rise to short range metallic and insulating charge ordered regions below T$_\mathrm{C}$, which order ferromagnetically and antiferromagnetically respectively. The localisation of the charge in the insulating regions, and the corresponding orbital order give rise to antiferromagnetism as understood through the context of the Goodenough Kanamori Anderson rules \cite{1963Goodenough}. The coexistence of the two electronically and magnetically ordered phases in the simple perovskite manganites is argued to be directly responsible for establishing colossal magnetoresistance \cite{1999Uehara, 2000Merithew, 2001Burgy, 2002Teresa, 2002Salamon}, although this has yet to be proven \cite{2007Shenoy, 2007Jiang}.

A similar picture is emerging in the columnar ordered quadruple perovskite manganites; competing exchange interactions and substitutional disorder are giving rise to a number of microscopically distinct magnetic phases. In the \tac (B = Mn) manganites the competing B-B and A-B exchange interactions have thus far given rise to well correlated magnetic structures on average. For instance in \ce{Tm2MnMnMn4O12} and \ce{Sm2MnMnMn_{4-x}Ti_xO12} a long range ordered collinear ferrimagnetic (FIM) phase was observed \cite{2019Vibhakar, 2020Vibhakar_SMTO}. It was suggested that the system ordered on average to favour the exchange interactions of the greatest strength, A-B exchange, while the competing exchange interactions, AFM B-B exchange, manifested as spin fluctuations reducing the measured moment. Similarly in \ce{$R$2CuMnMn4O12}($R$ = Y or Dy), well correlated magnetic phases were observed, however owing to a reduction in the A-B exchange interactions, B-B exchange was accommodated by spin canting onto the B-site layers \cite{2020Vibhakar_RCMO}. 

In this paper we show that the introduction of substitutional disorder on the A sites in the presence of the competing exchange interactions causes \ce{[Y_{2-x}Mn_x]MnMnMn4O12} $x=0.23$ and $x= 0.16$ to separate into two distinct magnetic phases; a collinear FIM phase that is favoured by A-B exchange, and an AFM phase favoured by B-B exchange. In analogy to the introduction of \emph{charge} doping in the simple perovskite manganites, the introduction of \emph{spin} doping through the substitution of magnetic Mn ions for non magnetic Y$^{3+}$, gives rise to a spin texture that is inhomogeneous on the nanoscale. In these systems we have observed the formation of short range AFM clusters contained in a paramagnetic matrix between T$_1$ and T$_2$. Below T$_2$, a phase transition to well correlated long range FIM order was observed, which coexists with the short range AFM clusters. A sharp increase in the AFM phase fraction was observed below $\sim$ 33 K, concomitant with a decrease in the FIM phase fraction. We note that characterising such effects of spin doping is of significance, given recent trends to replace established electronic paradigms with spin-only functionality in future spintronic devices. 

\section{Experimental Details} 

\ce{Y_{1.840(7)}Mn_{0.160(8)}MnMnMn4O12} ($x=0.16$) and \ce{Y_{1.768(9)}Mn_{0.232(9)}MnMnMn4O12} ($x=0.23$) powder samples were made from stoichiometric mixtures of \ce{Mn2O3} and \ce{Y2O3} (99.9\%) using a high-temperature high-pressure synthesis method as detailed in Ref. \cite{2017Zhang}. DC magnetization measurements were performed on the samples using a SQUID magnetometer (Quantum Design, MPMS-XL-7T) between 2 and 400 K under both zero-field-cooled (ZFC) and field-cooled on cooling (FCC) conditions. Time of flight neutron powder diffraction (NPD) measurements were performed on 1.93 g of the \Ypn sample and 1.51 g sample of the \Ypnf sample using the WISH diffractometer at ISIS \cite{2011ChaponWISH}. Data with high counting statistics were collected at 1.5 K and 85 K for both samples; temperatures representative of each of the magnetically ordered phases. Data with lower counting statistics were collected for both samples on warming between 1.5 K and 85 K in 2 K and 5 K steps, between 90 K and 140 K in 10 K steps, and at 200 K and 250 K. Small angle neutron scattering (SANS) measurements were performed on D33 at the Institute Laue Langevin (ILL). Each of the samples listed above were pressed into a flat disc like pellet, and placed in an aluminium sample mount. Data was collected in a single detector setting for both samples (0.003{\AA}$^{-1} \leqslant$ Q $\leqslant$ 0.1\AA$^{-1}$ $\Delta$Q = 0.0005 {\AA}$^{-1}$), and at temperatures between 300 K and 5 K. Field dependent SANS measurements were performed on the \Ypn sample with the magnetic field applied parallel to the incoming neutron beam, and data was collected at 5 K, 90 K and 130 K using fields of 0.01 T, 0.1 T and 0.5 T.

\begin{figure}[tph]
\centering
\includegraphics[width =\linewidth]{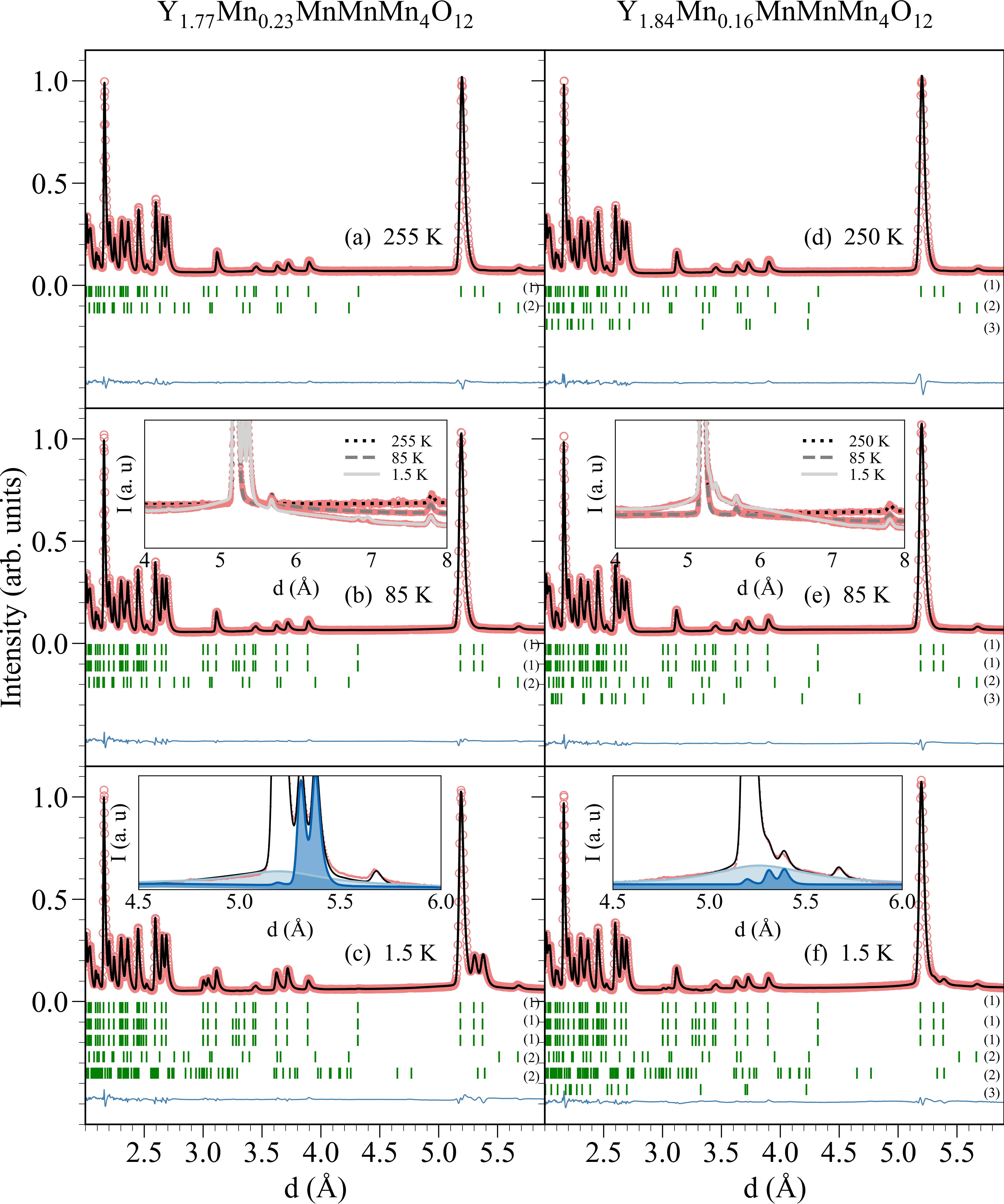}
\caption{\label{FIG::7_YMO_neutron}Neutron powder-diffraction data collected on the time-of-flight diffractometer WISH at ISIS at temperatures representative of the paramagnetic and two magnetically ordered phases of (a-c) $x=0.23$ and (d-f) $x=0.16$  \ce{[Y_{2-x}Mn_x]MnMnMn4O12}. The insets in (b) and (e) compare the NPD patterns at either 250 or 255 K, 85 K and 1.5 K for each of the samples. The insets in (c) and (f) show the contributions of the AFM phase and FIM phase to the \{110\} family of reflections, given by the light and dark blue shaded regions respectively for the diffraction pattern collected at 1.5 K. The data are represented by the red circles and the fit of the data by the black or grey lines. The green ticks marks represent the nuclear and magnetic reflections of the main and any impurity phases. The reflections labelled (1) are from the main \ce{[Y_{2-x}Mn_x]MnMnMn4O12} phase, (2) from the \ce{YMn2O5} impurity and (3) the \ce{YMnO3} impurity.}
\end{figure}

\section{Results} 

\subsection{\label{SEC::NPD}Neutron powder diffraction data}

\begin{table*}[tph]
\caption{\label{TAB::7_STRUCFAC}Reflection conditions for B-site $F_i$, $A_i$, $X_i$, $Y_i$ symmetry adapted basis modes as determined by structure factor calculations shown in Sec. \ref{SEC::STRUCFAC}. When two B-site modes are listed, the first mode corresponds to the magnetic ordering of the Mn3 layer, and the second mode to the magnetic ordering of the Mn4 layer. For instance the notation $A_iA_{\bar{i}}$ equates to the Mn3 ions transforming by an $A_i$ mode and the Mn4 ions transforming by an $A_{\bar{i}}$ mode, where the barred subscript implies the Mn4 sublattice is antialigned with respect to Mn3. $f_m$ is the magnetic form factor, which is approximated to be same for the Mn3 and Mn4 ions. Note that the direction of the magnetic moment has not been taken into account when calculating the structure factors.}
\setlength{\tabcolsep}{4.25pt}
\begin{ruledtabular}
{\renewcommand{\arraystretch}{1.5}
\begin{tabular}{c| c c c c c c c c c c c c c}
Mode  & (001)   & \{200\}   & (010) & (100)  &  (011)  & (101)  & (110)   &  (111) & (012) & (102)  & (112)    & (121)  & (211)\\
\hline
\hline
 & \multicolumn{13}{c}{A-sites}\\
$F_i$  & 0  & 2$f_m$ &  0   & 0      &  2$f_m$    & 2$f_m$  &  2$f_m$   & 0      &   0     & 0      &  2$f_m$ & 2$f_m$ & 2$f_m$ \\
$A_i$  &  2$f_m$  & 0 & 2$f_m$   & 2$f_m$       &   0     & 0      & 0 & 2$f_m$&   2$f_m$   &  2$f_m$     & 0  &  0     & 0\\
\hline
\hline
 & \multicolumn{13}{c}{B-sites}\\
$F_iF_i$         & 0       &  8$f_m$     & 0      &   0     & 0      & 0  &    0   &   0   &   0    &  0 &  0     & 0 & 0\\
$F_iF_{\bar{i}}$ & 8$f_m$  & 0      & 0      &   0     & 0      & 0       &0 &   0   &   0    &   0      &  0     & 0 & 0\\
$F_i$            & 4$f_m$  & 4$f_m$ &  0   & 0      &   0     & 0      &  0   & 0      &   0     & 0      & 0 & 0 & 0 \\
$A_iA_i$           & 0 & 0 &  0     & 0      &   0     & 0      & 8$f_m$  &    0   &   0   &   0    &  8$f_m$  &  0     & 0 \\
$A_iA_{\bar{i}}$   & 0 & 0 & 0   & 0      &   0     & 0      & 0       & 8$f_m$ &   0   &   0    &   0      &  0     & 0 \\
$A_i$             & 0  & 0 & 0  & 0      &   0     & 0      & 4$f_m$  & 4$f_m$ &   0   &   0    & 4$f_m$   &  0     & 0\\
$X_iX_i$     & 0  & 0 & 0    &  8$f_m$ &   0     & 0      & 0       & 0      & 0       & 8$f_m$ & 0   & 0      & 0 \\
$X_iX_{\bar{i}}$     & 0   & 0  & 0  & 0       &   0     & 8$f_m$ & 0       & 0      & 0       &   0    & 0   & 8$f_m$ & 0  \\
$X_i$              & 0 & 0  & 0     & 4$f_m$  &   0     & 4$f_m$ & 0       & 0      & 0       & 4$f_m$ & 0   & 4$f_m$ & 0 \\
$Y_iY_i$            &0& 0 &  8$f_m$  & 0       &   0         & 0      & 0       & 0      &  8$f_m$ &   0    & 0   & 0      & 0 \\
$Y_iY_{\bar{i}}$   & 0  &0  & 0   & 0       &   8$f_m$     & 0      & 0       & 0      &   0     &   0    & 0   & 0      & 8$f_m$ \\
$Y_i$   & 0 & 0 & 4$f_m$  & 0     & 4$f_m$       & 0      & 0       & 0 &  4$f_m$ &   0    & 0   & 0      & 4$f_m$\\
\end{tabular}
}
\end{ruledtabular}
\end{table*}

The published $Pmmn$ crystal structure model \cite{2017Zhang} was used to refine the NPD data collected in the paramagnetic phase, at 250 K for the \Ypnf sample and 255 K for the \Ypn sample, as shown in Fig. \ref{FIG::7_YMO_neutron}(a) and Fig. \ref{FIG::7_YMO_neutron}(b) respectively. The A sites (labelled Y1 and Y2) were predominately occupied by non magnetic Y$^{3+}$ and a small amount of magnetic Mn ions. The degree of cation disorder on these sites is what differentiates the two samples. As discussed later we show that it has profound implications for tuning exchange interactions, as it in effect leads to spin doping of the A sublattice \footnote{We cannot distinguish as to whether homovalent or heterovalent substitution occurs, i.e. if it is Mn$^{3+}$ or Mn$^{2+}$ that substitutes Y$^{3+}$ on the A sites. Hence it is possible that in addition to spin doping, charge doping may also occur. However as these samples remain insulating \cite{2017Zhang}, in comparison to the simple perovskite mangnaites where charge doping can induce a metal to insulator phase transition, we suggest that the changes we observe to the magnetic properties of these systems can largely be attributed to the effect of changing the spin properties of the A site ions.}. The \Ad sites were occupied by Mn$^{3+}$ in a square planar coordination (labelled Mn1) and the \Add sites by Mn$^{2+}$ in a tetrahedral coordination (labelled Mn2). The two symmetry inequivalent B sites, labelled Mn3 and Mn4, form $ab$ layers composed of Mn$^{3+}$ and Mn$^{3.5+}$ ions respectively, that stack alternately along $c$. The refined stoichiometries of the two \ce{Y2MnMnMn4O12} samples were \ce{Y_{1.768(9)}Mn_{0.232(9)}MnMnMn4O12} and \ce{Y_{1.840(7)}Mn_{0.160(8)}MnMnMn4O12}. Both samples contained a \ce{YMn2O5} impurity, with a weight fraction of 1.26 wt. \% and 1.32 wt. \% respectively. The \Ypnf sample also contained a \ce{YMnO3} impurity, with a weight fraction of 0.37 wt.\%. An excellent fit was achieved for both samples, $R = 3.30\%$, $wR = 3.30\%$, $R_\mathrm{Bragg} = 3.04\%$ for \Ypn at 250 K and $R = 3.36\%$, $wR = 3.71\%$, $R_\mathrm{Bragg} = 3.02\%$ for \Ypnf at 255 K. The refined crystal structure parameters are given in Table \ref{TAB::8_YMO_p78_xyz_bvs_PM} and Table \ref{TAB::8_YMO_p83_xyz_bvs_PM} of Sec. \ref{SEC::App_CryStruc} of the Appendix. 

As the temperature was lowered below 200 K, three distinct changes were observed to the NPD pattern of both samples. Firstly, between 200 K and 120 K the paramagnetic background gradually decreased, see inset of Fig. \ref{FIG::7_YMO_neutron}(b) and (e), but no additional intensity was observed across the rest of the patterns. Secondly, between 130 K and 110 K broad diffuse intensity started to appear centred about the \{110\} and \{112\} families of reflections, inset of Fig. \ref{FIG::7_YMO_neutron}(c) and (f), which steadily grew till $\sim$33 K, before rapidly increasing till the lowest measured temperatures. It was not possible to pinpoint the exact onset temperature of the broad diffuse magnetic intensity and therefore an approximate value of 120 $\pm$ 10 K will be given for T$_{1}$ in the rest of the paper. Thirdly, below T$_{2}$, which was 67 K for the \Ypn sample and 64 K for the \Ypnf sample, well correlated intensity appeared at the (011), (101), (121), (211), (020) and (200) Bragg reflections, Fig. \ref{FIG::7_YMO_neutron}. The intensity of these Bragg reflections was significantly higher for the \Ypn than the \Ypnf sample, and grew till $\sim$33 K, before steadily decreasing till the lowest measured temperatures.

The intensity that appeared below T$_{2}$ was identified to be magnetic in origin as the magnetic susceptibility measured on the sample underwent a rapid increase at T$_2$, indicative of the ordering of a FM or FIM phase. The broad diffuse intensity that was observed to appear below T$_{1}$ was also identified to have a magnetic origin given that it directly competes with the sharp well correlated magnetic intensity below 33 K. Details of whether a signature of a phase transition at T$_{1}$ is observed in magnetometry or specific heat capacity is discussed further in Sec. \ref{SEC::DC} and Sec. \ref{SEC::SPECHEAT} of the Appendix respectively.  

\begin{figure}
\centering
\includegraphics[width =\linewidth]{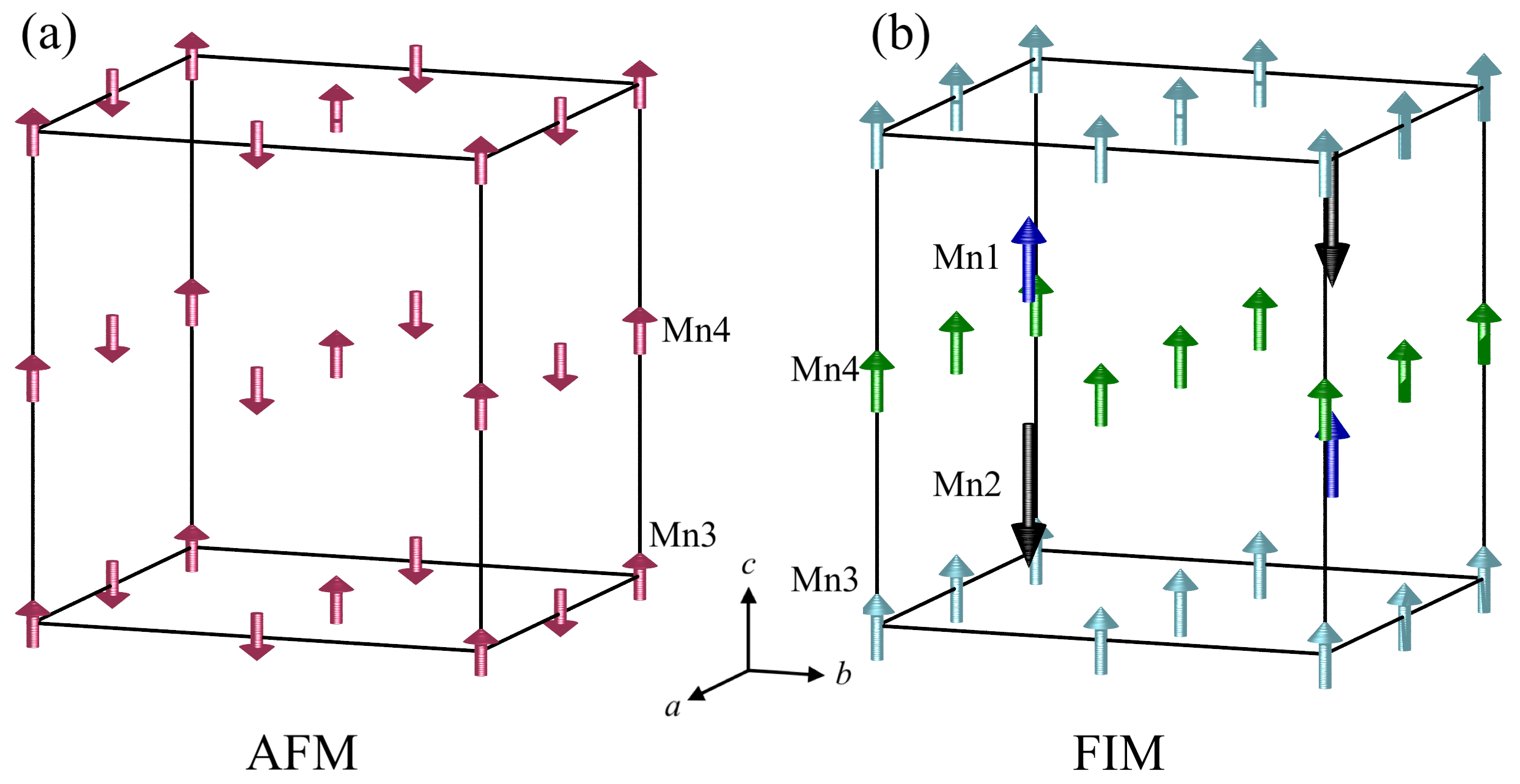}
\caption{\label{FIG::7_YMO_magstruc} Experimentally determined magnetic structure of (a) the AFM phase and (b) FIM phase of the \Ypn sample of \ce{[Y_{2-x}Mn_x]MnMnMn4O12} at 1.5 K.}
\end{figure}

As all the magnetic intensity was observed on top of structural reflections, the magnetic ordering was determined to transform by a $\Gamma$-point irreducible representation. The magnetic $\Gamma$-point representation decomposes into seven irreducible representations, determined using the \textsc{isotropy} software suite, and which are given in terms of the symmetry-adapted magnetic modes, $F_i$, $A_i$, $X_i$ and $Y_i$, defined in the Appendix, Sec. \ref{SEC::SYMANA}. Structure factors were calculated for each of the symmetry adapted basis modes, Sec. \ref{SEC::STRUCFAC} of the Appendix, and the results are summarised in Table \ref{TAB::7_STRUCFAC} for relevant reflections. These structure factor calculations show that the magnetic intensity observed below T$_{2}$ on the (020) and (200) reflections can originate from either A or B-site FM $F_i$ modes, Table \ref{TAB::7_STRUCFAC}. The observation of equal amounts of magnetic intensity on the (200) and (020), and zero magnetic intensity on the (002) reflections are consistent with the ordering of $F_z$ modes, which transform by the $\Gamma^+_2$ irrep. Magnetic structure models were constructed by taking linear combinations of the symmetry adapted basis functions of the $\Gamma_2^+$ irrep, and refined against the NPD data using the \textsc{fullprof} software suite. The magnetic structure model that gave the best fit to the sharp magnetic Bragg peaks at 40 K, where a smoothly varying background was used to fit the broad diffuse magnetic intensity, was one in which the Mn1, Mn2, Mn3 and Mn4 sublattices order with $F_z$ modes. The Mn2 sublattice moments were oriented antiparallel to the Mn1, Mn3 and Mn4 sublattice moments to give a net collinear FIM structure as shown in Fig. \ref{FIG::7_YMO_magstruc}; similar to both magnetically ordered phases of \ce{Tm2MnMnMn4O12} \cite{2019Vibhakar} and all three magnetically ordered phases of \ce{$R$2CuMnMn4O12} ($R$ = Y or Dy) \cite{2020Vibhakar_RCMO}. This magnetic structure solution is consistent with the rapid increase of the susceptibility measured below T$_{2}$, as discussed in further detail in Sec. \ref{SEC::DC}.

\begin{figure}
\centering
\includegraphics[width =\linewidth]{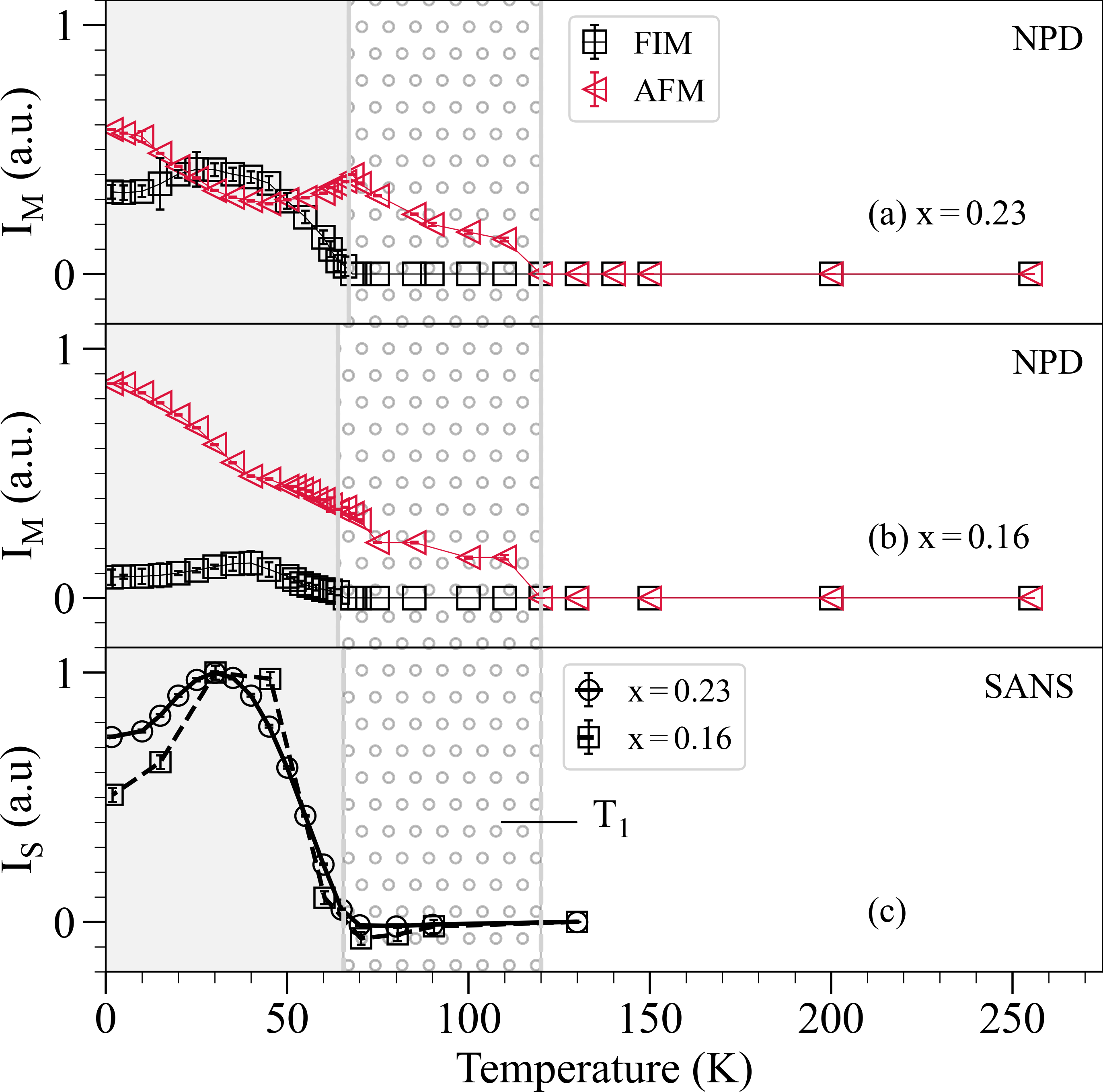}
\caption{\label{FIG::7_YMO_tempdep} I$_{\mathrm{M}}$ represents the average temperature evolution of magnetic scattering intensity from each phase for the (a) x = 0.23 and (b) x = 0.16 samples. I$_{\mathrm{M}}$ can be taken to indicate the evolution of the magnetic moments squared, scaled by their respective phase fraction. (c) Temperature dependence of the normalised magnetic SANS intensity for the \Ypn and \Ypnf samples.}
\end{figure}

The diffuse magnetic intensity centred about the \{110\} and \{112\} reflections, observed to appear below T$_{1}$, could have originated from AFM B-site modes or $F_i$ A-site modes coupled antiferromagnetically, see Table \ref{SEC::MAGMOD} and Sec. \ref{SEC::STRUCFAC} of the Appendix for further details. As these A and B-site modes originate in different irreps, it is unlikely that both sublattices contributed to the diffuse intensity, and owing to the broadness of the diffuse intensity it was impossible to distinguish which of these two sets of modes contributed to the NPD data. Instead SANS measurements confirmed the absence of any FM ordering between T$_1$ and T$_2$, as detailed in Sec. \ref{SEC::SANS}, confirming that the diffuse intensity originates in AFM modes at the B-sites. 

The AFM B-site modes that can scatter at the \{110\} and \{112\} reflections are the $A_iA_i$, $X_iX_{\bar{i}}$ and $Y_iY_{\bar{i}}$ modes \footnote{A single $A_i$ mode, i.e on a single B site layer, could also give rise to scattering on the (110) and (112), however significant intensity is also expected on the (111) reflection which was not observed.}, as shown in Table \ref{TAB::7_STRUCFAC}. Each of these three B-site AFM mode combinations was tested against the NPD data, where the amplitude of the modes on the Mn3 and Mn4 layer were constrained to be the same. The best fit to the diffuse intensity was using a magnetic structure model in which the B-site magnetic structure transforms by an $A_zA_z$ mode ($\Gamma_1^{+}$ irrep) to give an AFM structure, shown in Fig. \ref{FIG::7_YMO_magstruc}.

The above magnetic structure models were used to simultaneously fit the NPD data; the AFM structure was used to fit the broad diffuse scattering and the FIM structure to fit the sharp Bragg peaks. Since there is 100\% correlation between the moment size and the phase fraction, it was not possible to get these two quantities separately for the AFM and FIM phases without any additional constraints. The unconstrained refinement can only provide information about the moment size scaled by the respective phase fraction. Thus, the integrated intensities of the AFM and FIM magnetic reflections as a function of temperature also represent temperature evaluation of the scaled moments (squared) and not the absolute moment size as in the case of a single magnetic phase. To emphasise this fact, we call these intensities as average magnetic scattering intensities, I$_M$, and their temperature evolution is shown In Fig. \ref{FIG::7_YMO_tempdep}. In addition to the phase transitions discussed above, one can observe that below $\sim$33 K, the average magnetic intensity scattered from the FIM phase begins to decrease, which occurs concomitantly with an increase in the scattered intensity from the AFM phase. By assuming the B-site moments were of the same magnitude in the AFM and FIM phases at 1.5 K, the moment magnitudes of all relevant sublattices and phase fractions of the AFM and FIM phases were refined, as given in Table \ref{TAB::Magneticmoments}. The model gave a very good fit to the data, $R_{\mathrm{Mag}} = 3.04\%$ for the \Ypn sample and $R_{\mathrm{Mag}} = 7.39\%$ for the \Ypnf sample at 1.5 K. Note that the larger value of $R_{\mathrm{Mag}}$ for the \Ypnf sample is owing to the weak intensity that was scattered from the FIM phase. The phase fraction of the FIM phase was larger in the \Ypn sample compared with \Ypnf sample, consistent with our qualitative observations described above.

\begin{table}[tph]
\caption{\label{TAB::Magneticmoments}Moment magnitudes and phase fractions (P.F.) refined at  1.5 K for the $x = 0.23$ and $x = 0.16$ samples for the AFM and FIM phases for each of the symmetry inequivalent magnetic sublattices, under the assumption that the B-site moments are of the same magnitude in the AFM and FIM phases at 1.5 K.}
\setlength{\tabcolsep}{8pt}
\begin{ruledtabular}
{\renewcommand{\arraystretch}{1.2}
\begin{tabular}{c | c c c c c}
\multicolumn{6}{c}{$x = 0.23$} \\
\hline
\hline
    & Mn1     & Mn2      & Mn3    & Mn4 & P.F.\\ 
FIM & 2.5(2) & -4.0(2) & 1.8(3) & 1.7(3) & 0.4(2)\\
AFM & - & - & 1.8(3) & 1.7(3) & 0.6(2)\\

\hline
\hline
\multicolumn{6}{c}{$x = 0.16$} \\
\hline
\hline
    &  Mn1     & Mn2      & Mn3    & Mn4 & P.F.\\ 
FIM & 2.4(7) & -1.8(7) & 1.6(3) & 1.5(4) & 0.14(6) \\
AFM & - & - & 1.6(3) & 1.5(4) & 0.86(6) \\
\end{tabular}
}
\end{ruledtabular}
\end{table}

\subsection{\label{SEC::SANS}Small angle neutron scattering}

SANS measurements were performed to determine whether the broad diffuse scattering measured below T$_1$  originated from AFM or FM order. The magnetic scattering length density from SANS is given by the average magnetic moment in a unit cell, which is zero for AFM order, and hence the presence of finite SANS intensity is indicative of FM or FIM ordering. 

The transmitted intensity from the full beam and the scattered intensity from an empty sample cell was used to normalise the SANS patterns so that any contributions from the transmitted beam and aluminium sample holder were subtracted away. The SANS pattern collected at room temperature was also subtracted to remove the nuclear contribution to the scattered intensity. Fig. \ref{FIG::7_YMO_tempdep}(c) shows the variation in the magnetic SANS intensity summed over the collected Q range and plotted as a function of temperature for both samples. No SANS intensity was observed between T$_1$ and T$_2$ of either sample indicating the absence of any FM clusters in this temperature region. Furthermore in the presence of any FM clusters, one would expect the application of a magnetic field to couple to the net magnetisation of the cluster and cause a change to the magnetic SANS intensity, as has been observed in other systems \cite{2006Magen}. No change to the SANS intensity was observed between T$_1$ and T$_2$ under applied magnetic fields of upto 0.5 T, as shown in Fig. \ref{FIG::7_YMO_SANS_F} of the Appendix, confirming the absence of any FM order.

The magnetic SANS intensity that begins to appear below T$_2$ in both samples, Fig. \ref{FIG::7_YMO_tempdep}(c), reproduces the the observed variation in intensity of the Bragg peaks measured from NPD. Hence, we assign this SANS intensity to the onset of the FIM phase, which is expected to scatter at the (0,0,0) reflection, the tail of which is observed in the SANS data.

\begin{figure*}[ht]
\centering\includegraphics[width =\linewidth]{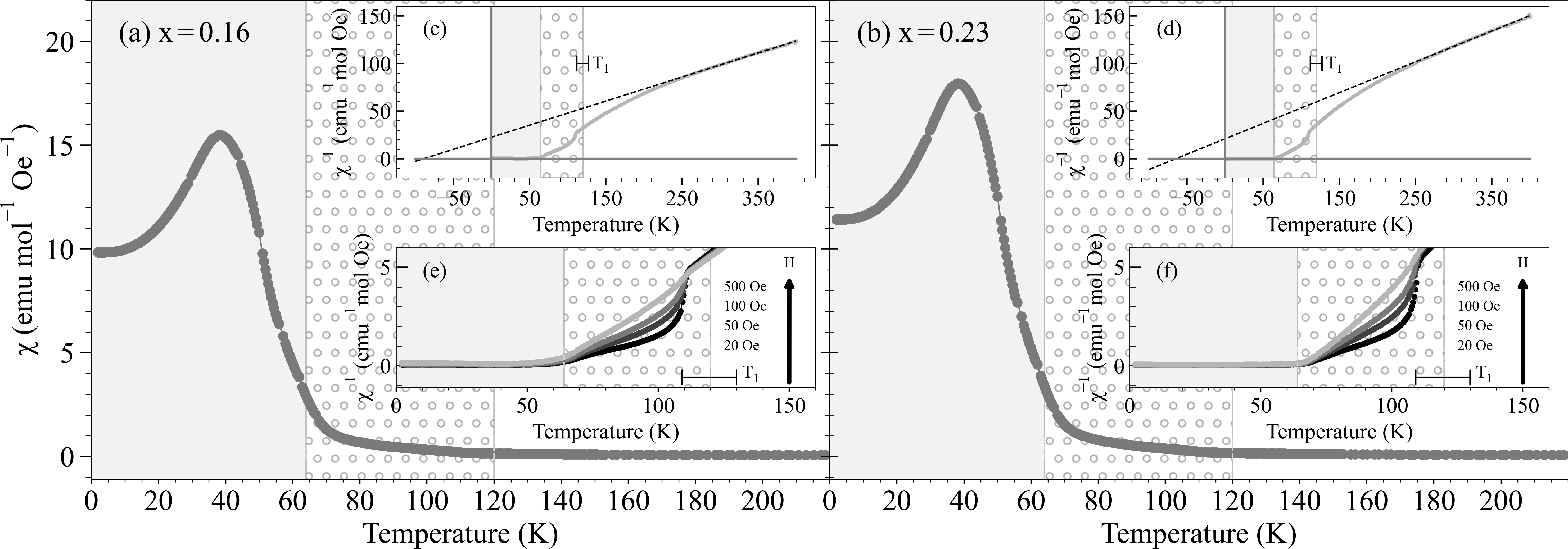}
\caption{\label{FIG::7_YMO_magnetometry} Temperature dependence of the FCC susceptibility under an applied field of 100 Oe for (a) the $x = 0.16$ sample (b) the $x = 0.23$ sample. FCC inverse susceptibility measurements collected under 100 Oe field with a Curie Weiss fit of the 400 - 250 K data shown by the dotted line for (c) the $x = 0.16$ sample (d) the $x = 0.23$ sample. FCC inverse susceptibility measurements under different applied DC fields for (e) the $x = 0.16$ sample (f) the $x = 0.23$ sample.}
\end{figure*} 

\subsection{\label{SEC::DC}DC susceptibility measurements}

Between 400 K and 250 K the inverse susceptibility follows a Curie Weiss like temperature dependence, as shown in the insets (c) and (d) of Fig. \ref{FIG::7_YMO_magnetometry}, with a Curie Weiss temperature of $-65$ K for the \Ypn sample and $-89$ K for \Ypnf sample. Between 200 K and T$_{1}$ a departure from Curie Weiss like behaviour was observed, there was a slight curvature to the inverse susceptibility. Taken together with the decrease in the paramagnetic background of the NPD patterns at these temperatures, suggests the onset of short range magnetic correlations.

A downturn in the inverse susceptibility was observed between T$_{2}$ and 110 K, shown in insets (e) and (f) of Fig. \ref{FIG::7_YMO_magnetometry}, and which is suppressed upon application of a magnetic field. This is widely regarded in the literature to be characteristic of a Griffiths phase consisting of ferromagnetic clusters \cite{2002Salamon, 2006Magen, 2006Ouyang}. However, since our experimental results indicate that only antiferromagnetic clusters are present in this temperature region, which intuitively cannot cause a downturn to the inverse susceptibility, this feature is likely due to the presence of a FM or FIM impurity phase. The high-temperature high-pressure synthesis of the \tacm manganites are susceptible to the formation of impurities, such as the orthorhombic \ce{R_{0.667}Mn_{0.333}MnO_3} manganites, which have ordering temperatures between 105 and 110 K and are expected to order ferromagnetically or ferrimagnetically \cite{2018Zhang}. While there are no diffraction peaks in the NPD pattern that would be consistent with a \ce{Y_{0.67}Mn_{0.33}MnO3} impurity, we note that magnetometry measurements can be more sensitive to the presence of such impurities than  diffraction. Below T$_2$ the magnetic susceptibility rapidly increases, which is consistent with the onset of the long range FIM ordering. 

\subsection{Discussion}

In the simple perovskite manganites, \ce{AMnO3}, the magnetic structure immediately below T$_\mathrm{N}$ is determined by the superexchange interactions between Mn ions at the B-sites; we refer to these as B-B interactions. The Goodenough Kanamori Anderson rules then provide a framework in which one can predict the sign and approximate strength of the B-B exchange, depending upon the orbital configuration of the Mn-O-Mn superexchange pathways. Applying the same rules to the B-B exchange interactions in the columnar ordered quadruple perovskites one predicts an AFM B site structure composed of $A_iA_i$ modes. However, in all previously studied columnar ordered perovskites \cite{2019Vibhakar, 2020Vibhakar_RCMO, 2020Vibhakar_SMTO} ferromagnetic modes are found on the B sites, and together with the A sites these materials adopt a ferrimagnetic structure. A key difference between the simple and quadruple perovskites is the presence of magnetic transition metal ions at the A-sites (A{\textquotesingle} and A{\textquotesingle}{\textquotesingle} sites in the perfectly ordered columnar systems), and it was suggested that ferromagnetism at the B-sites in the columnar ordered stystems, is stabilised by A-B exchange interactions dominating B-B interactions. This frustration between A-B and B-B exchange has manifested in reduced ordered moments \cite{2020Vibhakar_SMTO} and large spin canting \cite{2020Vibhakar_RCMO}.  

Here, in \ce{[Y_{2-x}Mn_x]MnMnMn4O12}, we have observed the first example of an AFM $A_iA_i$ B-site magnetic structure in the quadruple perovskites, which emerges as a separate, poorly correlated phase that competes with the typical ferrimagnetic long-range order. Note that we can rule out the scenario of spin canting within a single phase as both F and A modes are polarised in the same direction ($z$). The AFM phase appears at a higher temperature than the FIM phase, likely due to the effects of frustration in the FIM case. This is consistent with symmetry; the A-sites do not enter into the $\Gamma_1^+$ irreducible representation by which the B-site $AzAz$ modes transform. Hence, for $AzAz$ ordering at the B-sites the A-B exchange interactions exactly cancel by symmetry. To the contrary, a net A-B exchange is allowed within the $\Gamma_2^+$ symmetry of the ferrimagnetic phase, which will compete with the B-B interactions.

We propose that the observed phase separation is driven by inhomogeneities in the sample, such that in minority regions B-B exchange wins over A-B interactions allowing B-site AFM clusters to develop. In the $x = 0.23$ sample 11.5(5)\% of the Y sites are occupied by Mn, and in the x$ = 0.16$ sample 8.0(4)\% of the Y sites are occupied by Mn. This local substitution of non-magnetic Y by magnetic Mn can be considered \emph{spin doping} of the A-site sublattice, which will tune the competition between A-B and B-B exchange in favour of the FIM structure. Inhomogeneities in the amount of spin doping can then give rise to AFM / FIM phase separation, whereby clusters of antiferromagnetic order form in regions of the crystal where there is minimal substitutional disorder on the A-sites. This scenario naturally explains the larger AFM phase fraction found for $x=0.16$ (less spin doping) compared to $x=0.23$ (more spin doping), as shown in Fig. \ref{FIG::7_YMO_tempdep}.

Finally, we comment on the suppression of the FIM phase in favour of the AFM phase at low temperature, an effect that is enhanced in the lower spin-doped $x=0.16$ sample. We assume that the changes in average magnetic intensity observed in the NPD experiments reflect a change in phase fraction, as a reduction in the FIM moment on cooling would be unphysical for this insulating system. The observed changes in the average intensity scattered from the AFM / FIM phases, reflective of the changing phase fractions, implies that the average competition between A-B and B-B exchange has been tuned by temperature. The ratio of A-B to B-B exchange will change in line with the changing ratio of the A and B site magnetic moments, such that at a certain temperature, regions of the crystal with a critical concentration of spin doping could switch from being FIM to AFM.

\section{Conclusions}

In conclusion we have found that the \ce{[Y_{2-x}Mn_x]MnMnMn4O12} $x=0.23$ and $x=0.16$ quadruple perovskite manganites developed short range antiferromagnetic clusters below T$_{1}$  = 120 $\pm$ 10 K. Below T$_{2}$ $\sim$ 65 K the system was composed of well correlated long range collinear ferrimagnetic order, punctuated by short range antiferromagnetic clusters. Furthermore we observed a sharp increase in the antiferromagnetic phase fraction below $\sim$ 33 K, concomitant with a decrease in the ferrimagnetic phase fraction. We propose that substitutional disorder on the A-sites caused an inhomogeneous distribution of exchange interactions across the sample, such that in some regions of the crystal where magnetic Mn replaces non-magnetic Y$^{3+}$ (increasing A-B exchange) the FIM phase was favoured to grow, while in other regions, where non-magnetic Y$^{3+}$ is retained on the A-sites (minimising A-B exchange), B-B exchange dominated and the AFM phase was stabilised. Our results show that in the \tacm manganites less than 10\% spin doping, introduced through cation disorder on the A sites, can have a profound impact on the magnetic phases stabilised in the presence of competing exchange between multiple magnetic sublattices. In future work one might anticipate forming Griffiths type phases linked to emergent material properties such as colossal magnetoresistance through control of the spin degrees of freedom.

\section{Appendix}

\subsection{\label{SEC::App_CryStruc}Crystal Structure Parameters}

In the \Ypn sample 13.2(4)\% of Y1 sites and 10.0(4)\% of Y2 sites were occupied by Mn ions, and in the \Ypnf sample 10.0(4)\% of Y1 sites and 6.0(4)\% of Y2 sites were occupied by Mn ions. Both samples contained a \ce{YMn2O5} impurity, with a weight fraction of 1.26 wt. \% and 1.32 wt. \% respectively. The \Ypnf sample also contained a \ce{YMnO3} impurity, with a weight fraction of 0.37 wt.\%. An excellent fit was achieved for both samples, $R = 3.30\%$, $wR = 3.30\%$, $R_\mathrm{Bragg} = 3.04\%$ for the \Ypn sample at 250 K and $R = 3.36\%$, $wR = 3.71\%$, $R_\mathrm{Bragg} = 3.02\%$ for the \Ypnf sample at 255 K. The refined crystal structure parameters for each sample are given in Tables \ref{TAB::8_YMO_p78_xyz_bvs_PM} and \ref{TAB::8_YMO_p83_xyz_bvs_PM}. 

\clearpage
\begin{table*}[tph]
\setlength{\tabcolsep}{0.95em}
\caption{\label{TAB::8_YMO_p78_xyz_bvs_PM}Crystal structure parameters of \Ypn ($Z=4$, space group $Pmmn$) refined at 255 K. The lattice parameters were determined to be $a =  7.24742(7) ~ \mathrm{\AA}$, $b = 7.43400(2) ~ \mathrm{\AA}$, and $c = 7.78894(9) ~ \mathrm{\AA}$. Excellent reliability parameters of $R = 3.30\%$, $wR = 3.30\%$, $R_\mathrm{Bragg} = 3.04\%$ were achieved in the refinement. Bond valence sums (BVS) were calculated using the bond valence parameters, $R_0$(Y$^{3+}$) = 2.01(1), $R_0$(Mn$^{3+}$) = 1.76(1), $R_0$(Mn$^{2+}$) = 1.79(1), $R_0$(Mn$^{3+}$) = 1.76(1), $R_0$(Mn$^{4+}$) = 1.75(1), B = 0.37 \cite{1991Brese}.}
\begin{ruledtabular}
{\renewcommand{\arraystretch}{1.4}
\begin{tabular}{c | c c c c c c c}
Atom & Site & $x$ & $y$ & $z$ & $U_\mathrm{iso}$ $(\mathrm{\AA}^2)$ & B.v.s. ($|e|$) & Occ. \\
\hline
Y1  & $2a$ & 0.25 & 0.25 & 0.7816(4) & 0.0067(1) & 2.92(5)& 0.217(1)\\
Y2  & $2a$ & 0.25 & 0.25 & 0.2839(4) & 0.011(2) & 2.88(8) & 0.225(1)\\
Mn1  & $2b$ & 0.75 & 0.25 & 0.7227(6) & 0.002(1) & 2.70(5) & 0.25\\
Mn2  & $2b$ & 0.75 & 0.25 & 0.2393(8) & 0.019(5) & 1.96(6) & 0.25\\
Mn3  & $4c$ & 0 & 0 & 0 & 0.002(1) & 3.16(9) & 0.5\\
Mn4  & $4d$ & 0 & 0 & 0.5 &  0.09(1) & 3.6(1) & 0.5\\
O1   & $8g$ &  0.4349(2) & -0.0682(2) &  0.2673(2) & 0.0111(6) & - & 1.0 \\
O2   & $4f$ & 0.0568(3) &  0.25 &  0.0422(3)  & 0.0105(7) & - & 0.5 \\
O3   & $4e$ & 0.25 & 0.5296(3) & 0.9220(3) & 0.0087(7) & - & 0.5 \\
O4   & $4f$ & 0.5417(3) &  0.25 & 0.4159(3) & 0.0109(7)  & - & 0.5 \\
O5   & $4e$ & 0.25 & 0.4338(3) &  0.5403(3) &  0.0095(7) & -  & 0.5\\
\end{tabular}
}
\end{ruledtabular}
\end{table*}

\begin{table*}[tph]
\setlength{\tabcolsep}{0.95em}
\caption{\label{TAB::8_YMO_p83_xyz_bvs_PM}Crystal structure parameters of \Ypnf ($Z=4$, space group $Pmmn$) refined at 250 K. The lattice parameters were determined to be $a =  7.25284(8) ~ \mathrm{\AA}$, $b = 7.44842(8) ~ \mathrm{\AA}$, and $c = 7.7963(1) ~ \mathrm{\AA}$. Excellent reliability parameters of $R = 3.49\%$, $wR = 3.79\%$, $R_\mathrm{Bragg} = 3.14\%$ were achieved in the refinement. Bond valence sums (BVS) were calculated using the bond valence parameters, $R_0$(Y$^{3+}$) = 2.01(1), $R_0$(Mn$^{3+}$) = 1.76(1), $R_0$(Mn$^{2+}$) = 1.79(1), $R_0$(Mn$^{3+}$) = 1.76(1), $R_0$(Mn$^{4+}$) = 1.75(1), B = 0.37 \cite{1991Brese}.}
\begin{ruledtabular}
{\renewcommand{\arraystretch}{1.4}
\begin{tabular}{c | c c c c c c c c}
Atom & Site & $x$ & $y$ & $z$ & $U_\mathrm{iso}$ $(\mathrm{\AA}^2)$ & B.v.s. ($|e|$) & Occ.\\
\hline
Y1  & $2a$ & 0.25 & 0.25 & 0.7814(4) & 0.0120(9) & 2.91(5) & 0.226(1)\\
Y2  & $2a$ & 0.25 & 0.25 & 0.284(4) & 0.0081(9) & 2.89(5) & 0.235(1)\\
Mn1  & $2b$ & 0.75 & 0.25 & 0.7246(7) & 0.004(1) & 2.66(5) & 0.25\\
Mn2  & $2b$ & 0.75 & 0.25 & 0.2402(1) & 0.033(2) & 1.93(6) & 0.25 \\
Mn3  & $4c$ & 0 & 0 & 0 & 0.0076(1) & 3.2(1) & 0.5\\
Mn4  & $4d$ & 0 & 0 & 0.5 &  0.0074(1) & 3.5(1) & 0.5\\
O1   & $8g$ & 0.4352(2) & -0.0677(2) & 0.2667(2) & 0.0126(6) & - & 1.0\\
O2   & $4f$ & 0.0570(3) & 0.25 & 0.0415(3) & 0.0111(7) & - & 0.5\\
O3   & $4e$ & 0.25 & 0.5291(3) & 0.9223(3) &  0.0089(8)  & - & 0.5\\
O4   & $4f$ & 0.5401(3) &  0.25 &  0.4148(3)  & 0.0126(8) &- & 0.5\\
O5   & $4e$ & 0.25 & 0.4334(3) & 0.5399(3) & 0.0097(7) & - & 0.5\\
\end{tabular}
}
\end{ruledtabular}
\end{table*}

\section{\label{SEC::MAGSYM}Magnetic Symmetry}
\subsection{\label{SEC::MAGMOD}Magnetic Modes}

Symmetry-adapted modes that constitute candidate magnetic structures for all ordered phases of \ce{[Y_{2-x}Mn_x]MnMnMn4O12} are given below. We define four modes, $F_i$, $A_i$, $X_i$, and $Y_i$, that span the four symmetry equivalent B site Mn ions of a given Mn3 ($z=0$) or Mn4 ($z=1/2$) layer in a single $Pmmn$ unit cell (the subscript $i = x,~y,~\mathrm{or}~z$ denotes the direction of the moment component, where $x||a$, $y||b$, and $z||c$), adopting the same labelling scheme as established for other members of the columnar ordered quadruple perovskites \cite{2020Vibhakar_RCMO}. The magnetic order of a single layer is then described by a linear combination of these modes. For each mode the relative signs of the magnetic moment components on each of the four sites are given in Table \ref{TAB::Bmagneticmodes}. We note that under the physical constraint that every symmetry equivalent magnetic ion has the same size magnetic moment, combined modes must have orthogonal magnetic moment components. 

\begin{table*}[tph]
\caption{\label{TAB::Bmagneticmodes}Magnetic modes that span the four symmetry equivalent B site Mn ions of a Mn3 ($z=0$) or Mn4 ($z=1/2$) layer in a single $Pmmn$ unit cell. The subscript $i$ denotes the direction of the moment component.}
\begin{ruledtabular}
{\renewcommand{\arraystretch}{1.2}
\begin{tabular}{c | c  c  c  c} 
Frac. coords.  & $F_{i}$ & $A_{i}$ & $X_{i}$ & $Y_{i}$\\
\hline
0.0, 0,0, $z$ & + & + & + &  + \\
0.5, 0.0, $z$ & + & - & - &  + \\
0.0, 0.5, $z$ & + & - & + &  - \\
0.5, 0.5, $z$ & + & + & - &  - \\
\end{tabular}
}
\end{ruledtabular}
\end{table*}

\begin{table*}[tph]
\caption{\label{TAB::Amagneticmodes}Magnetic modes that span the two symmetry equivalent A site ions of a $R$1, $R$2, Cu1, or Mn2 sublattice in a single $Pmmn$ unit cell. The subscript $i$ denotes the direction of the moment component.}
\begin{ruledtabular}
{\renewcommand{\arraystretch}{1.2}
\begin{tabular}{c | c  c c} 
Frac. coords.  & $F_i$ & $A_i$ & \\
\hline
$x$, $y$, $z$ & + & + & \\
$x+1/2$, $y+1/2$, $z+1/2$ & + & - & \\
\end{tabular}
}
\end{ruledtabular}
\end{table*}

\begin{table*}[tph]
\caption{\label{TAB::3_orth_irreps} The magnetic $\Gamma$-point irreducible representations described for each of the symmetry inequivalent magnetic sites of the \ce{[Y_{2-x}Mn_x]MnMnMn4O12} $Pmmn$ \tac crystal structure using the symmetry adapted basis modes defined above.}
\setlength{\tabcolsep}{18pt}
\begin{ruledtabular}
{\renewcommand{\arraystretch}{1.2}
\begin{tabular}{c|cccccc} 
Irrep. & Y1 & Y2  & Mn1 & Mn2 & Mn3 & Mn4 \\
\hline
\multicolumn{7}{c}{$\mathrm{k}=(0,0,0)$} \\
\hline
$\Gamma_1^+$ & - & - & - & - & $Y_x,X_y,A_z$ & $Y_x,X_y,A_z$ \\
$\Gamma_2^+$ & $F_z$ & $F_z$ & $F_z$ & $F_z$ & $X_x,Y_y,F_z$ & $X_x,Y_y,F_z$ \\
$\Gamma_3^+$ & $F_x$ & $F_x$ & $F_x$ & $F_x$ & $F_x,A_y,X_z$ & $F_x,A_y,X_z$ \\
$\Gamma_4^+$ & $F_y$ & $F_y$ & $F_y$ & $F_y$ & $A_x,F_y,Y_z$ & $A_x,F_y,Y_z$ \\
$\Gamma_1^-$ & $A_z$ & $A_z$ & $A_z$ & $A_z$ & - & - \\
$\Gamma_3^-$ & $A_y$ & $A_y$ & $A_y$ & $A_y$ & - & - \\
$\Gamma_4^-$ & $A_x$ & $A_x$ & $A_x$ & $A_x$ & - & - \\ 
\end{tabular}
}
\end{ruledtabular}
\end{table*}

As for the A sites, we define two modes, $F_i$ and $A_i$, that span the two symmetry equivalent A site ions of a $R$1, $R$2, Mn1, or Mn2 sublattice in a single $Pmmn$ unit cell. The relative signs of the magnetic moment components on each of the two sites are given in Table \ref{TAB::Amagneticmodes}.

\subsection{\label{SEC::SYMANA}Symmetry Analysis}

The NPD experiments described in Section \ref{SEC::NPD} demonstrate a $\Gamma$-point propagation vector for all the  magnetically ordered phases of \ce{[Y_{2-x}Mn_x]MnMnMn4O12}. Symmetry analysis was performed using \textsc{isodistort} \cite{2006Isodistort}, taking the $Pmmn$ structure detailed in Section \ref{SEC::App_CryStruc} as the parent. The magnetic $\Gamma$-representation for the Wyckoff positions describing magnetic sites of the \ce{[Y_{2-x}Mn_x]MnMnMn4O12} $Pmmn$ crystal structure each decompose into seven irreducible representations, listed in Table \ref{TAB::3_orth_irreps}, and described in terms of the symmetry adapted basis modes defined above.

\subsection{\label{SEC::STRUCFAC}Structure Factor Calculations}

To simplify the structure factor calculations of the A-sites, they are approximated to have a $z$ fractional coordinates of exactly 1/4 or 3/4. From these calculations, Eqns. \ref{EQN::SFC_A_Fi} and \ref{EQN::SFC_A_Ai}, one can see that the $F_i$ mode has finite intensity for reflections where $h + k + l$ is even and zero intensity for reflections where $h + k + l$ is odd. In contrast the $A_i$ mode has finite intensity for reflections where $h + k + l$ is odd, and zero intensity for reflections where $h + k + l$ is even. The presence of broad diffuse intensity about the \{110\} and \{112\} families of reflections, and the absence of any broad diffuse intensity about the \{2,0,0\} family is only consistent with the Mn1 and Mn2 sublattices ordering with $F_i$ modes where the Mn1 and Mn2 sublattices are coupled antiferromagnetically.  In the FIM phase the Mn1 and Mn2 sublattices do order with $F_i$ modes that are coupled antiferromagnetically, hence it is unlikely that this could be a competing magnetic phase to the FIM order. Furthermore if the broad diffuse intensity originated in $F_i$ A-site modes coupled antiferromagnetically, one may expect that as the long range FIM order develops, the broad diffuse intensity should shift into the sharp magnetic Bragg peaks. However the broad diffuse intensity \emph{grows} below T$_2$, and at the expense of the sharp magnetic Bragg peaks below $\sim$ 33 K.

\begin{widetext}

\begin{center}
\underline{A-sites}
\end{center}

\underline{$F_i$}
\begin{equation}
F_{(h, k, 1)}  = f_{\mathrm{Mn1}}(\exp^{\pi i k} + \exp^{\pi i (h + l)}) +  f_{\mathrm{Mn2}}(\exp^{\pi i (h)} + \exp^{\pi i(k+l)})
\label{EQN::SFC_A_Fi}
\end{equation}

\underline{$A_i$}
\begin{equation}
F_{(h, k, 1)}  = f_{\mathrm{Mn1}}(\exp^{\pi i k} - \exp^{\pi i (h + l)}) +  f_{\mathrm{Mn2}}(\exp^{\pi i (h)} - \exp^{\pi i(k+l)})
\label{EQN::SFC_A_Ai}
\end{equation}

\begin{center}
\underline{B-sites}
\end{center}

The structure factors equations for each of the symmetry adapted B-site ion basis modes is given by Eqns. \ref{EQN::SFC_B_Fi} - \ref{EQN::SFC_B_Yi}. Structure factor equations taking into account two B-site magnetic modes that describe the magnetic structure of the B-sites on the $z = 0$ layer (the first mode listed) and the $z = 1/2$ layer (the second mode listed) is given by Eqns. \ref{EQN::SFC_B_FiFi}-\ref{EQN::SFC_B_YiYi}, where the moment magnitudes across the two sublattices are constrained to be the same, and they are assumed to have the same form factor for simplicity.

\underline{$F_i$}
\begin{equation}
\label{EQN::SFC_B_Fi}
F_{(h, k, l)} = f_m ( 1 +  \exp^{\pi i h} +  \exp^{\pi i k} + \exp^{\pi i(h+k)} )
\end{equation}

\underline{$A_i$}
\begin{equation}
F_{(h, k, l)} = f_m ( 1 - \exp^{\pi i h} - \exp^{\pi i k} + \exp^{\pi i(h+k)} )
\end{equation}

\underline{$X_i$}
\begin{equation}
F_{(h, k, l)} = f_m ( 1 - \exp^{\pi i h} + \exp^{\pi i k} - \exp^{\pi i(h+k)} )
\end{equation}

\underline{$Y_i$}
\begin{equation}
\label{EQN::SFC_B_Yi}
F_{(h, k, l)} = f_m ( 1 + \exp^{\pi i h} - \exp^{\pi i k} - \exp^{\pi i(h+k)} )
\end{equation}

\underline{$F_iF_i$}
\begin{equation}
\label{EQN::SFC_B_FiFi}
F_{(h, k, l)} = f_m ( 1 + \exp^{\pi i h} + \exp^{\pi i k} + \exp^{\pi i(h+k)} + \exp^{\pi i l} + \exp^{\pi i (h + l)} + \exp^{\pi i (k + l)} + \exp^{\pi i(h + k  + l)})
\end{equation}

\underline{$F_iF_{\bar{i}}$}
\begin{equation}
F_{(h, k, l)} = f_m ( 1 + \exp^{\pi i h} + \exp^{\pi i k} + \exp^{\pi i(h+k)} -\exp^{\pi i l} - \exp^{\pi i (h + l)} - \exp^{\pi i (k + l)} - \exp^{\pi i(h + k  + l)})
\end{equation}

\underline{$A_iA_i$}
\begin{equation}
F_{(h, k, l)} = f_m ( 1 - \exp^{\pi i h} - \exp^{\pi i k} + \exp^{\pi i(h+k)} + \exp^{\pi i l} - \exp^{\pi i (h + l)} - \exp^{\pi i (k + l)} + \exp^{\pi i(h + k  + l)})
\end{equation}

\underline{$A_iA_{\bar{i}}$}
\begin{equation}
F_{(h, k, l)} = f_m ( 1 - \exp^{\pi i h} - \exp^{\pi i k} + \exp^{\pi i(h+k)} -\exp^{\pi i l}+ \exp^{\pi i (h + l)} + \exp^{\pi i (k + l)} - \exp^{\pi i(h + k  + l)})
\end{equation}

\underline{$X_iX_i$}
\begin{equation}
F_{(h, k, l)} = f_m ( 1 - \exp^{\pi i h} + \exp^{\pi i k} - \exp^{\pi i(h+k)} + \exp^{\pi i l} - \exp^{\pi i (h + l)} + \exp^{\pi i (k + l)} - \exp^{\pi i(h + k  + l)})
\end{equation}

\underline{$X_iX_{\bar{i}}$}
\begin{equation}
F_{(h, k, l)} = f_m ( 1 - \exp^{\pi i h} + \exp^{\pi i k} - \exp^{\pi i(h+k)} -\exp^{\pi i l} + \exp^{\pi i (h + l)} - \exp^{\pi i (k + l)} + \exp^{\pi i(h + k  + l)})
\end{equation}

\underline{$Y_iY_i$}
\begin{equation}
F_{(h, k, l)} = f_m ( 1 + \exp^{\pi i h} - \exp^{\pi i k} - \exp^{\pi i(h+k)} + \exp^{\pi i l} + \exp^{\pi i (h + l)} - \exp^{\pi i (k + l)} - \exp^{\pi i(h + k  + l)})
\end{equation}

\underline{$Y_iY_{\bar{i}}$}
\begin{equation}
\label{EQN::SFC_B_YiYi}
F_{(h, k, l)} = f_m ( 1 + \exp^{\pi i h} - \exp^{\pi i k} - \exp^{\pi i(h+k)} -\exp^{\pi i l} - \exp^{\pi i (h + l)} + \exp^{\pi i (k + l)} + \exp^{\pi i(h + k  + l)})
\end{equation}
\end{widetext}

\subsection{\label{SEC::SPECHEAT}SANS}

\begin{figure}[ht]
\centering
\includegraphics[width =\linewidth]{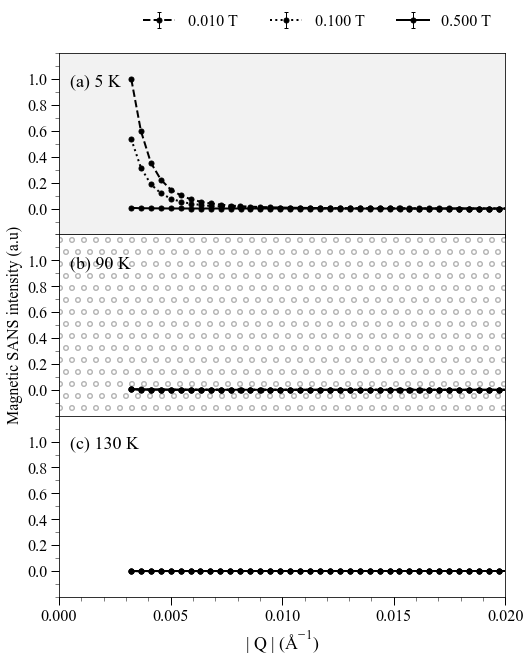}
\caption{\label{FIG::7_YMO_SANS_F} Normalised magnetic SANS intensity under different applied magnetic fields at (a) 5 K (b) 90 K and (c) 130 K for $x = 0.23$ sample.}
\end{figure}

Magnetic SANS intensity collected under applied magnetic fields of 0.01 T, 0.1 T and 0.5 T at 5 K, 90 K and 130 K for the x $ = 0.23$ sample are show in Fig. \ref{FIG::7_YMO_SANS_F}.

\subsection{Specific heat capacity measurements}

\begin{figure}[ht]
\centering
\includegraphics[width =\linewidth]{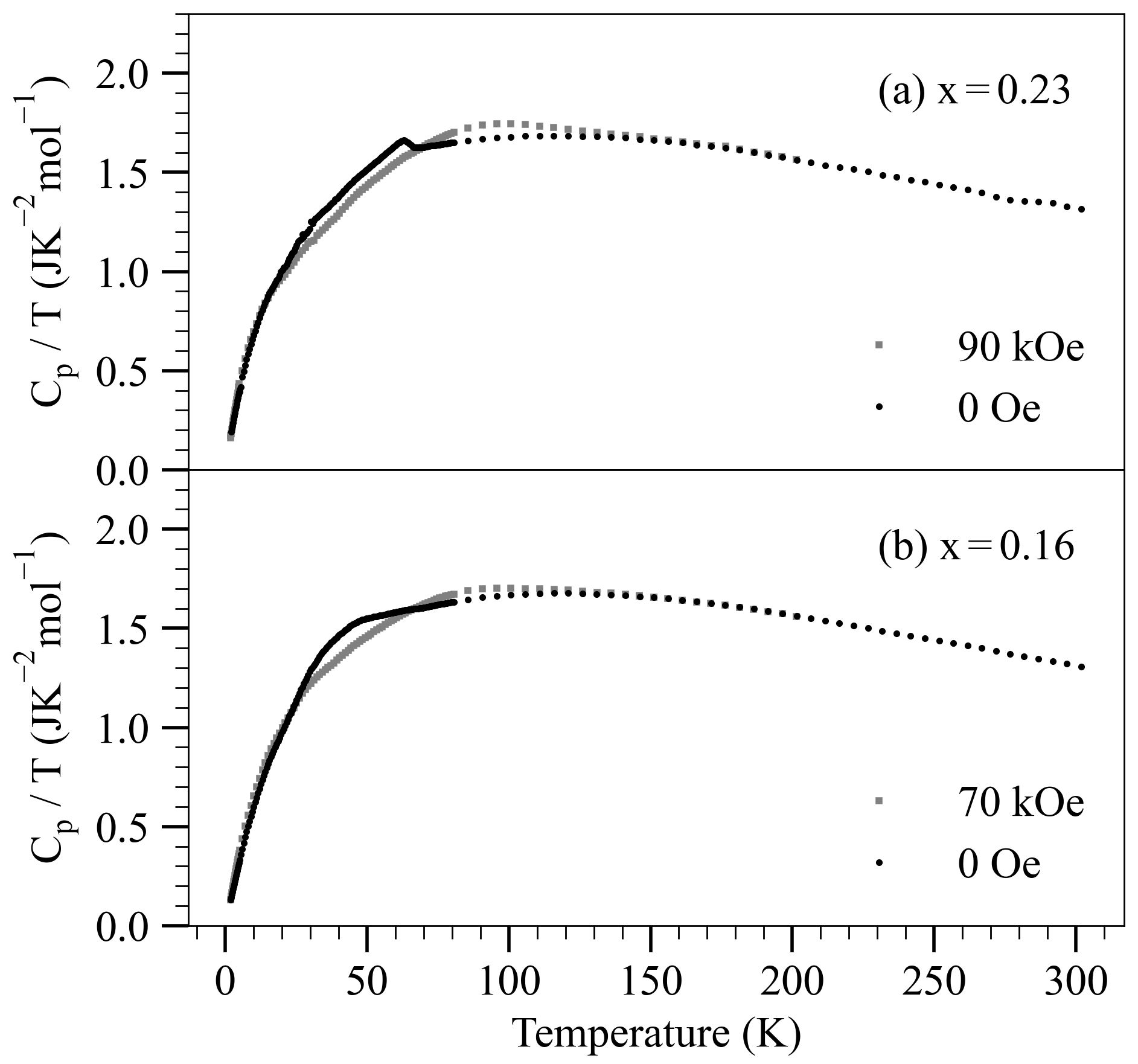}
\caption{\label{FIG::7_YMO_SPECIFICHEAT} Specific heat data shown for the (a) $x = 0.23$ and (b) the $x = 0.16$ \ce{[Y_{2-x}Mn_x]MnMnMn4O12} samples.}
\end{figure}

Specific heat data measured on both samples at two different fields is shown in Fig. \ref{FIG::7_YMO_SPECIFICHEAT}. Only a single phase transition, characteristic of the ordering of the FIM phase is observed in both samples.

\bibliographystyle{apsrev4-2}

\end{document}